\documentclass[prl,showpacs,twocolumn,aps]{revtex4}%
\usepackage{graphicx}
\usepackage{epsfig}
\usepackage[english]{babel}
\usepackage{amsmath}
\usepackage{amsfonts}
\usepackage{amssymb}%
\begin{document}
\title{Dynamic phase transitions in electromigration-induced step bunching}
\author{Vladislav Popkov}
\thanks{E-mail: popkov@thp.uni-koeln.de}
\author{Joachim Krug}
\thanks{E-mail: krug@thp.uni-koeln.de}
\affiliation{Institut f\"{u}r Theoretische Physik, Universit\"{a}t zu K\"{o}ln, Germany.}

\begin{abstract}
Electromigration-induced step bunching in the presence of sublimation or
deposition is studied theoretically in the attachment-limited
regime. We predict a phase transition as a function of the
relative strength of kinetic asymmetry and
step drift. For weak asymmetry the number of steps between bunches
grows logarithmically with bunch size, whereas for strong asymmetry at most a
single step crosses between two bunches.  
In the latter phase the emission and
absorption of steps is a collective process which sets in only above 
a critical bunch size and/or step interaction strength.
\end{abstract}
\date{\today }

\maketitle

Much of the morphological structure and dynamics of
crystal surfaces can be understood in terms of the behavior of steps that
separate different exposed atomic layers \cite{Jeong99,Krug05a}. Since they
entail a finite free energy cost per unit length, steps are long-lived
structural defects which nevertheless, due to their one-dimensional nature,
are highly sensitive to thermal fluctuations. These fluctuations induce
long-ranged steric interactions between steps, which complement similar
interactions mediated by bulk elasticity. When such a system of interacting
steps is driven out of equilibrium by external forces, e.g. during growth or
sublimation of the crystal surface, a rich variety of morphological patterns
and dynamic phenomena emerge.

As was first shown by Latyshev and coworkers \cite{Latyshev89}, step patterns
on Si(111) surfaces can be efficiently manipulated by a direct heating
current, which induces mass transport along the surface through the
electromigration of adatoms. Subsequently a multitude of
electromigration-generated step patterns have been found and studied
experimentally \cite{Yagi01,Minoda03}, including step bunches
\cite{Latyshev89,Yang96,Metois99,Fujita99}, step
antibands \cite{Thurmer99}, in-phase wandering steps
\cite{Degawa99} and step pairs \cite{Pierre-Louis04}, 
many of which still defy a comprehensive theoretical description. 

In this Letter we report on the surprising discovery of a novel type of phase
transition in the most basic model of electromigration-induced step bunching
originally introduced by Stoyanov \cite{Stoyanov91,Liu98,Stoyanov98,Sato99}. 
In this model the steps
are assumed to be straight, and the uniform step train is destabilized by an
electromigration force in the downhill direction. The phase transition occurs
as a function of a dimensionless parameter $b$, defined in (\ref{b_asymmetry})
below, which gauges the relative importance of electromigration-induced
kinetic asymmetry and step drift due to sublimation or growth. This parameter
can be tuned experimentally, e.g., by changing the electromigration force
through the DC component of the heating current, or the sublimation rate
through a change in temperature.%

\begin{figure}
[ptb]
\begin{center}
\includegraphics[
height=2.4in,
width=3.5in
]%
{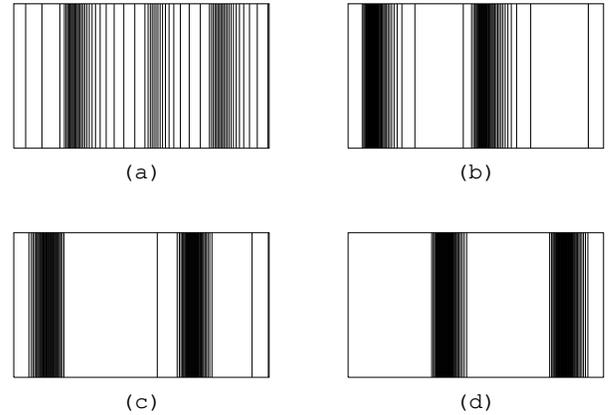}%
\caption{Typical step configurations (top view of a vicinal surface)
\ generated by numerical solution of (\ref{discrete_time_evolution}). 
Graphs (a)-(d) correspond to
$b=0.1,0.5,5,20$ respectively, and in all cases $U/bl=0.6$. Each frame
contains about 60 steps. Reduction of 
the number of free flowing steps with increasing $b$
is evident. In case (d) there are no free-flowing steps since the bunch
sizes are within the dead zone region of Fig.~\ref{Fig_phaseb10}}%
\label{Fig_multiplot_impulses}%
\end{center}
\end{figure}
%EndExpansion

Step drift leads to the exchange of steps between bunches, which plays an
important role in the evolution and coarsening of the bunch pattern \cite{Stoyanov98,Sato99}. 
The most striking visual signature of the phase transition is a qualitative change in
the number of such free steps, and in the mechanism by which they are
exchanged (Fig.~\ref{Fig_multiplot_impulses}). For $b < 1$ (strong drift/ weak
asymmetry) the step density decreases smoothly in the outflow region of a
bunch, and the number of steps between bunches grows logarithmically with the
bunch distance. In contrast, for $b > 1$ (weak drift/strong asymmetry) there
is at most a single free step between any two bunches, irrespective of their
size. This feature should make the two regimes clearly distinguishable in
experiments using reflection electron microscopy 
\cite{Latyshev89,Metois99} or scanning tunneling microscopy 
\cite{Yang96,Fujita99}.

The dynamics in the regime $b>1$ is remarkably complex. The exchange of a step
is a collective process involving both the expelling and the receiving bunch,
which sets in only beyond a critical bunch size, and which is
accompanied by breathing oscillations of the entire bunch. As a consequence, a
stationary bunch shape amenable to a continuum description \cite{Pimpinelli02}
of the type developed previously for $b<1$ \cite{Krug05b,Popkov05} does not
appear to exist.

\textit{Model.} We consider a system of straight, non-transparent steps
subject to electromigration and sublimation (including also a growth flux is
straightforward). We work in the attachment-detachment limited regime, where
the kinetic length $d=D/k$, the ratio of surface diffusion coefficient $D$ and
attachment rate $k$, is large compared to the step spacing
\cite{Krug05a}. The equations of
motion for the step positions $x_{i}(t)$ then take the form \cite{Liu98}
\[
\frac{dx_{i}}{dt}=\frac{1-b}{2}\left(  x_{i+1}-x_{i}\right)  +\frac{1+b}%
{2}\left(  x_{i}-x_{i-1}\right)
\]%
\begin{equation}
+U\left(  2f_{i}-f_{i-1}-f_{i+1}\right)  \label{discrete_time_evolution}%
\end{equation}
where the time scale has been normalized to the sublimation flux. 
Summing over $i$ we see that the average step velocity $v$ is equal to the 
mean terrace width $l$. In numerical solutions of (\ref{discrete_time_evolution}) 
lengths are measured in units of $l$, i.e. we set $l=1$.
The last term on the right hand side represents
stabilizing step-step interactions of strength $U$, where, for combined
entropic or dipolar elastic repulsion,
\begin{equation}
f_{i}=\left(  \frac{l}{x_{i}-x_{i-1}}\right)  ^{\nu+1}-\left(  \frac
{l}{x_{i+1}-x_{i}}\right)  ^{\nu+1}, \label{step-step interaction}%
\end{equation}
with $\nu=2$ \cite{Jeong99}. The
parameter $b$ governs the asymmetry between ascending and descending steps,
relative to the mean step velocity, which induces step bunching when $b>0$.
Linear stability analysis of (\ref{discrete_time_evolution}) shows that the
instability sets in at wavelengths corresponding to bunches
containing more than $M^{\ast}$ steps,
with%
\begin{equation}
M^{\ast}=2\pi [\arccos(  1- bl/12 U)]  ^{-1}.
\label{linear_stability_analysis}%
\end{equation}
In previous work more complicated
variants of the step equations (\ref{discrete_time_evolution}) have been studied
numerically, and some of the features analyzed in this paper have been described
on a qualitative level \cite{Stoyanov98,Sato99}. The advantage of using 
the attachment/detachment limited dynamics (\ref{discrete_time_evolution})
lies in the linearity of the destabilizing terms,
which allows to clearly expose the key role of
the parameter $b$ and the existence of a sharp phase transition.

In terms of physical quantities, the parameters $b$ and $U$ are given by
\cite{Liu98}%
\begin{equation}
b=\frac{\Gamma F\tau_{e}}{2k_{B}Ta^{2}},\;\;\;\;\;U=\frac{\Gamma\tau_{e}%
g}{2k_{B}T}\tan^{3}\alpha\label{b_asymmetry}%
\end{equation}
where $\Gamma$ is the step mobility for the Brownian motion of an isolated
step \cite{Jeong99}, $a^{2}$ is the atomic area, $F$ is the electromigration
force acting on an adatom, $\tau_{e}$ is the inverse desorption rate,
$\alpha=a/l$ is the miscut angle, and $g$ is the step interaction parameter.

The model (\ref{discrete_time_evolution}) is expected to apply in two of the
four temperature regimes \cite{Yagi01} in which step bunching is observed on
Si(111), around 900$^{\text{o}}$~C and around 1250$^{\text{o}}$~C
\cite{Metois99,Fujita99}. The parameters given in \cite{Yang96,Liu98}
lead to the estimates $b\approx14$ in the low temperature regime and
$b\approx0.3$ in the high temperature regime, which shows that both cases $b <
1$ and $b > 1$ are experimentally realizable. 

Step equations of the form (\ref{discrete_time_evolution}) can also be derived
for step bunching induced by Ehrlich-Schwoebel (ES) barriers during
sublimation \cite{Krug05b} or by inverse ES barriers during growth
\cite{Krug05b,Slanina05}. In this sense (\ref{discrete_time_evolution})
constitutes a rather generic model of step bunching kinetics. However, in step
bunching induced by ES barriers the parameter $b$ is restricted to the
interval $0 < b < 1$, and hence the phenomena described in this paper do not occur.

\textit{Structure of the outflow region.} In the presence of step drift,
coarsening of step bunches is a very dynamic process during which 
steps continuously leave (flow out of) one bunch
and join (flow into) its neighbour \cite{Stoyanov98,Sato99}. In \cite{Popkov05,Slanina05} it was shown
that the analysis of the outflow region provides key insights into the shape
and dynamics of bunches for $b<1$. We shall see now that there are drastic
differences between the outflow regions for the cases $b<1$ and $b>1$. We
consider a bunch containing a large number $M\gg1$ of steps, so that its shape can
be considered quasi-stationary. We impose
periodic boundary conditions $\Delta_{i}(t)=$ $\Delta_{i+M}(t)$ for the
terrace sizes $\Delta_{i}=x_{i+1}-x_{i}$. Stationarity implies then periodicity of 
each step trajectory (up to an overall shift with velocity $v=l$), with some period
$\tau(b,U,M)$, during which each step $i$ will once cross the plateau
between two consecutive bunches. After time $\tau/M$, each step $i$ will
substitute the position of step $i+1$ (up to a constant shift independent of $i$), so that%
\begin{equation}
\Delta_{i\pm1}(t)=\Delta_{i}(t\pm\frac{\tau}{M}). 
\label{delta(t+tau/M)}%
\end{equation}
Deriving an equation for $\Delta_{i}(t)$ from (\ref{discrete_time_evolution})
and substituting (\ref{delta(t+tau/M)}), we get a differential-difference 
equation for a single, $\tau$-periodic function $\Delta(t)=\Delta_{i}(t)$ \cite{Popkov05} 
\begin{equation}
\frac{d\Delta(t)}{dt}=\frac{1-b}{2}\Delta(t+\frac{\tau}{M})+b\Delta
(t)-\frac{1+b}{2}\Delta(t-\frac{\tau}{M})+U(...),\label{continuous_for_delta}%
\end{equation} 
where for brevity the $U$-containing
terms are only sketched. The (unknown)
period $\tau$ determines the
velocity of a bunch:  after time $\tau$ the bunch shifts by ($-Ml$)
in a frame co-moving with velocity $v$; in the laboratory frame 
its lateral velocity is then \cite{Popkov05} 
\begin{equation}
\label{V}
V=l(1-M/\tau).
\end{equation} 
Big bunches are separated by wide plateaux, and for the 
steps crossing a plateau (in case there are many) the $U$-term 
in (\ref{continuous_for_delta}) should
become negligible. In this outflow region one can solve the remaining linear
part by the ansatz $\Delta(t)\sim\exp(qMt/\tau)$ obtaining the transcendental
equation $b\left(  \cosh\left(  q\right)  -1\right)  =\sinh\left(  q\right)
-qM/\tau$. To fix the unknown parameter $\tau$, we recall the Fourier analysis
of (\ref{discrete_time_evolution}) in \cite{Popkov05}, which shows,
irrespective of the value of $b$, that for large $M$ 
generically $\tau(M)\approx M+O(1)$. Thus
\begin{equation}
b\left(\cosh q - 1 \right)  = \sinh q - q \label{b(q)}%
\end{equation}
which has a real positive solution for each $b<1$ but no solutions for
$b>1$. In the following we explore the consequences of this fact.

\textit{Number of steps between bunches.} For $b<1$, the existence of a
solution $q$ of (\ref{b(q)}) implies a smooth decrease of the step density in
the outflow region, with the terrace widths increasing exponentially, as
$\Delta_{k}/\Delta_{k-1}\approx\exp\left(  q\right)  $. To estimate the number
$N_{f}$ of free steps between two bunches of size $M$, we
equate the total length $\sim l\exp[qN_{f}]$ occupied by $N_{f}$ terraces to
the typical distance $Ml$ between the bunches, and obtain
$
N_{f}\approx q^{-1}\ln M.
$
For small $b$, the solution of (\ref{b(q)}) can be
approximated by $q\approx3b$.

For $b\rightarrow1$, $q$ diverges and $N_{f}$ vanishes. 
The absence of solutions of (\ref{b(q)})
means that the $U$-term in (\ref{discrete_time_evolution}) can never be
neglected and that correspondingly there can be at most one step crossing the
plateau between two bunches, at any stage of evolution. One can check, using
(\ref{discrete_time_evolution}) and (\ref{step-step interaction}), that any
configuration with more than one step between two bunches is unstable
for $b>1$, so that all steps except at most one will be pushed back to the
bunch they originated from. In Fig.~\ref{Fig_multiplot_impulses} we 
show numerically generated 
bunch configurations in the course of coarsening for $b<1$ and $b>1$,
which confirm this conclusion.%

\begin{figure}
[ptb]
\begin{center}
\includegraphics[
height=2.2in,
width=3.3in
]%
{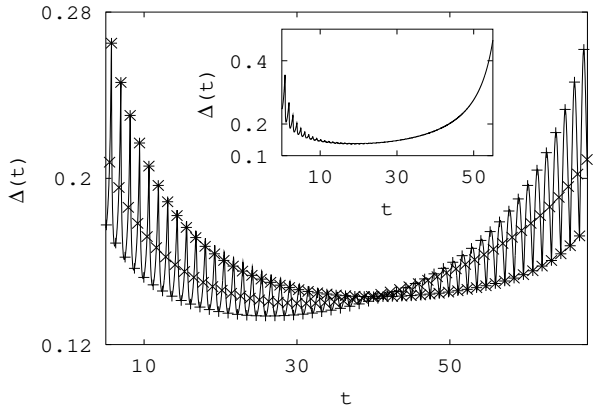}%
\caption{Full line: 
Distance $\Delta(t)$ between a pair of neighboring steps moving
through the bunch interior (initial and end regions excluded),
for $b=10,U/l=6,M=64.$ Symbols show the sizes of $42$
consecutive terraces (out of $64$), at times $t=0$ (+),
$t = 0.34\cdot(\tau/M)$ ($\times$), $t\approx0.55\cdot(\tau/M)$ ($\ast$),
illustrating the oscillatory breathing of the bunch. They
lie on the curve $\Delta(t)$ because of (\ref{delta(t+tau/M)}).
\textbf{Inset:} $\Delta(t)$ for $b=0.176$, $U/l=0.108$. Oscillations 
are triggered by a step colliding with the front end of the bunch, but do not
extend into the outflow region. }%
\label{Fig_Delta_inset}%
\end{center}
\end{figure}
%EndExpansion

\textit{Dynamics of emission and absorption of steps.} We now examine in more
detail how steps are emitted from a bunch. It is seen directly from
(\ref{discrete_time_evolution}) that the last step (with label $i$, say) of
the bunch at position $x_{i}$, which is trailing a wide terrace of width
$\Delta_{i}=x_{i+1}-x_{i}\gg l$, will be driven to escape from the bunch by
the linear term $(1-b)\Delta_{i}/2$ , provided $b<1$. This term indeed gives
the main contribution to the dynamics of the last step of the bunch, as we see
from numerical analysis. The emitted step does not perturb the remaining
steps; the $(i-1)$-th step which has become the last, is free to escape once
the $i$-th step has travelled sufficiently far. Bunches emit steps
continuously, creating an outflow region governed entirely by the linear part
of (\ref{discrete_time_evolution}) \cite{Popkov05,Slanina05}.

In contrast, for $b>1$, the linear term $(1-b)\Delta_{i}/2$ in
(\ref{discrete_time_evolution}) gives a negative contribution to the step
velocity, and the only way to move the last step $i$ away from the bunch is by
step-step interactions [the $U$-term in (\ref{discrete_time_evolution})]. 
Since the next step $i-1$ cannot be emitted before step $i$ has
landed at the next bunch, the configuration of steps at the
end of the bunch has to be changed by the emission process -- if it were
unchanged, the next step $i-1$ would be emitted immediately after the $i$-th.
This gives rise to oscillations of the bunch profile at the end, which spread
to the whole bunch, and whose amplitude grows with increasing $b$. Such
oscillations at the outflow end of the bunch are completely absent in the
$b<1$ phase (Fig.~\ref{Fig_Delta_inset}).

\begin{figure}
[ptb]
\begin{center}
\includegraphics[
height=2.2in,
width=3.4in
]%
{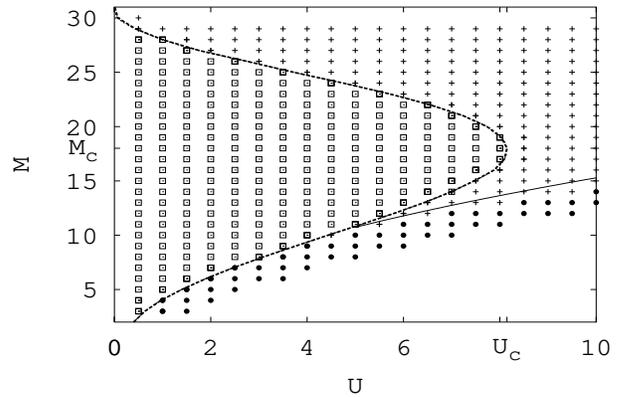}%
\caption{Phase diagram characterising quasi-stationary step
bunches of size $M$ as function of step-step repulsion $U$ for $b=10$, 
$l=1$. Three
phases can be defined: (a) Bunches smaller than the size
$M^{\ast}$ deduced from linear stability analysis
[Eq.(\ref{linear_stability_analysis})] dissolve (filled circles). (b) Dead
zone: for $M_{c1}(U)<M<M_{c2}(U)$ bunches are stable but do not
emit steps (squares). (c) For $M>M_{c2}(U)$ and $M^{\ast}<M<M_{c1}(U)$
bunches are stable and emit steps (crosses). The lines $M_{c1}(U),M_{c2}(U)$
terminate at a critical interaction strength $U_{c}(b)$ beyond which stable
bunches always emit steps. 
}%
\label{Fig_phaseb10}%
\end{center}
\end{figure}

When the emitted step finally collides with the receiving
bunch it provokes perturbations in the inflow region of the bunch, which are
visible both for $b<1$ and $b>1$. In the case $b<1$, however, the oscillations
in the inflow region are damped and disappear towards the interior of the
bunch. On the contrary, in the $b>1$ phase the oscillations penetrate through
the bunch, regain their large amplitude towards the bunch tail, and culminate
in the emission of the last step of the bunch, provided that the initial
impact was sufficiently strong (Fig.~\ref{Fig_Delta_inset}). 
The persistence of oscillations through the
bunch interior implies correlations between the emission and absorption of
steps, which should have important consequences for the coarsening dynamics;
this question will be addressed elsewhere. 

\begin{figure}
[ptb]
\begin{center}
\includegraphics[
height=2.3in,
width=3.5in
]%
{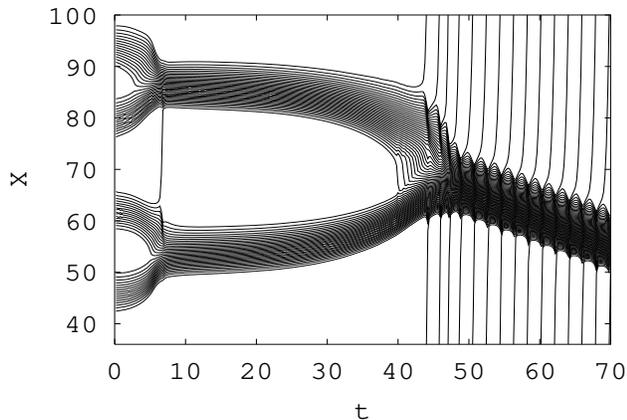}%
\caption{Space-time plot of individual step trajectories during coarsening,
in a frame co-moving with the step velocity $v$, for $b=20,U/l=12$. 
Step emission starts when the bunch size exceeds 
a critical value. The isolated emission event around $t \approx 7$ occurs
because the bunch shapes are not yet stationary. The initial
configuration is a slightly perturbed train of bunches consisting of $16$ steps
each.}%
\label{Fig_coarsening_b20}%
\end{center}
\end{figure}

\textit{Onset of step emission.} We have seen above that the emisssion of
steps in the $b>1$ regime is a nontrivial dynamical phenomenon facilitated by
a large step-step repulsion $U$, and suppressed by the kinetic asymmetry $b$.
For small $U$ (or  large $b$) the oscillatory breathing of the
bunch may not be able to trigger the emission of steps when  bunches are
small. The typical behavior of bunches as a function of  size $M$ and step
interaction $U$ at a fixed value of $b>1$ is summarized in the phase diagram
in Fig.~\ref{Fig_phaseb10}. For any given $b>1$, there exists a critical value
$U_{c}(b)$ such that for $U<U_{c}$ bunches emit steps only for sizes
$M^\ast < M < M_{c1}$ and $M > M_{c2}$, whereas for $U>U_{c}$ stable bunches 
always emit steps. Inside the dead zone $M_{c1} < M < M_{c2}$ the time 
interval $\tau/M$ between emission of steps is infinite, and correspondingly
bunches move with the mean step speed, $V = v$ [see (\ref{V})]. 
The ratio $\tau/M$ decays
monotonically to $1$ with distance from the dead zone, 
reflecting the fact that $\lim_{M \to \infty} \tau/M = 1$ for 
any fixed $b$ \cite{Popkov05}.

Diagrams for different $b$ can be superimposed 
after rescaling $U_{c}$ and $M_{c} \equiv 
M_{c1,2}(U_c)$ according to the relations
$ U_{c}\approx0.0105 \cdot b^{\alpha} $ and $ M_{c}\approx 2.112 \cdot b^{\gamma}$
with $\alpha\approx2.87$, $\gamma\approx0.935$ for all parameters investigated
($3<b\leq25$, $10^{-2}\leq U\leq120$), with a relative error not
exceeding $5\%$. Note that the relation $\gamma=(\alpha-1)/2$ implies
invariance of the linear instability curve (\ref{linear_stability_analysis})
at large $U$ under rescaling.

Different step kinetics for bunches of different sizes implies a change in
coarsening dynamics, highlighted in Fig.~\ref{Fig_coarsening_b20}. For 
$b\gg1$, depending on the value of $U$ different
coarsening scenarios are possible. For $U>U_{c}$ steps are exchanged
throughout the coarsening process, while for $U\lesssim
U_{c}/2$ late stage coarsening proceeds in two stages: without step exchange
(for bunches sizes smaller than $M_{c2}$) and with step emission once the
typical bunch size exceeds $M_{c2}$. 
Coarsening with or without step exchange has previously been associated
with nonconserved ($b$ finite) and conserved ($b = \infty$, no sublimation) dynamics,
respectively \cite{Sato99}; here we see that both types of behavior
may coexist when $b > 1$.

\textit{Conclusions.} We predict a new type of phase transition in
electromigration-induced step bunching within the regime of nontransparent
steps and attachment-detachment limited kinetics. The transition is
characterized by a dramatic change in the number and behavior of the free
steps that are exchanged between bunches, which should be clearly visible in
experiments on surfaces vicinal to Si(111). Theoretical challenges for the
future include the development of a continuum description for $b > 1$, and the
investigation of the correlated coarsening dynamics in this regime.

This work has been supported by DFG within project KR 1123/1-2.

\bibliographystyle{apsrev}
% \bibliography{popkov}

\end{document}